\documentstyle[12pt]{article}

\def\k2{\kappa^2}
\def\km2{\kappa^{-2}}
\def\lm{\lambda}
\def\sg{\sigma}

\def\Gm{\Gamma}
\def\om{\omega}

\def\pd{\partial}
\def\hf{{1 \over 2}}

\def\dl{\delta}

\def\hg{{\hat g}}
\def\hR{{\hat R}}

\begin{document}
\thispagestyle{empty}
\begin{flushright}
OU-HET-455 \\ LTH 605\\ CERN-TH/2003-245
\end{flushright}
\vskip30pt
\begin{center}
{\Large {\bf Wheeler-DeWitt Equation \\ in AdS/CFT
Correspondence}}
\end{center}
\vskip25pt
\begin{center}
{\large Takahiro Kubota$^\dag$, Tatsuya Ueno$^\ddag$ and Naoto
Yokoi$^\sharp$

\vskip20pt

$^\dag$\, Graduate School of Science,

Osaka University,

Toyonaka, Osaka 560-0043, Japan

{\it kubota@het.phys.sci.osaka-u.ac.jp}

\vskip20pt

$^\ddag$\, Department of Mathematical Sciences

University of Liverpool

Liverpool, L69 3BX, England

{\it ueno@liv.ac.uk}

\vskip20pt

$^\sharp$\, Theory Division, CERN

CH-1211 Geneva 23, Switzerland

{\it Naoto.Yokoi@cern.ch}}

\end{center}

\vskip30pt

\begin{center}
{\bf ABSTRACT}
\end{center}
We discuss a quantum extension of the holographic RG flow equation
obtained previously from the classical Hamiltonian constraint in
the bulk AdS supergravity. The Wheeler-DeWitt equation is proposed
to generate the extended RG flow and to produce $1/N$ subleading
corrections systematically. Our formulation in five dimensions is
applied to the derivation of the Weyl anomaly of boundary ${\cal
N}=4$ $SU(N)$ super-Yang-Mills theory beyond the large $N$ limit.
It is shown that subleading $1/N^2$ corrections arising from
fields in ${\rm AdS}_5$ supergravity agree with those obtained
recently by Mansfield et al. using their Schr\"{o}dinger equation,
thereby guaranteeing to reproduce the exact form of the boundary
Weyl anomaly after summing up all of the KK modes.

\pagebreak \setcounter{page}{1}

Since the proposal of the AdS/CFT correspondence by Maldacena
\cite{m}, a lot of effort has been made to test his conjecture.
Among several others, Henningson and Skenderis derived the
boundary theory Weyl anomaly in the large $N$ limit by evaluating
the tree-level on-shell action of the bulk AdS supergravity
\cite{hs}. (See also \cite{no}.) The same result was also obtained
by Mansfield and Nolland \cite{mn}, where they constructed a
functional Schr\"{o}dinger equation for the partition function on
AdS space and solved it at the tree-level. Recently the
Schr\"{o}dinger equation was fully considered to obtain a
subleading correction to the large $N$ result, and  the exact form
of the boundary Weyl anomaly has been successfully derived
\cite{mn2},\cite{mnu}.
\par

On the other hand, an alternative formulation of the bulk theory
was given by de Boer, Verlinde and Verlinde \cite{dbvv}, who gave
attention to the fact that the radial flow in AdS space transverse
to boundary directions corresponds to the renormalization group
(RG) flow at the boundary. The classical Hamiltonian constraint
arising from the reparameterization invariance of the bulk
supergravity with respect to (w.r.t.) the radial direction was
shown to be cast into the Callan-Symanzik RG flow equation on the
boundary space. In this formulation, the derivation of the
anomaly, at least in the large $N$ limit, was seen to be performed
more simply and quickly than with the other methods.
\par

The purpose of this article is to extend the classical Hamiltonian
constraint to a quantum equation on the bulk AdS and to present a
generalized holographic RG flow equation, which systematically
produces subleading corrections in the $1/N$ expansion. Obviously,
the Wheeler-DeWitt equation is a promising candidate for it and
suggests that the subleading correction arises from terms with
second-order functional derivatives w.r.t. fields of AdS
supergravity, as happens in the ordinary WKB method of quantum
mechanics. We demonstrate that the Wheeler-DeWitt equation applied
to the $\rm AdS_5/CFT_4$ case works indeed to derive the exact
$1/N^2$ correction to the Weyl anomaly of the boundary ${\cal
N}=4$ $SU(N)$ super-Yang-Mills (SYM) theory.
\par

\vspace{1.2cm}

We start with a brief review of the tree-level (large $N$)
calculation of the boundary Weyl anomaly via the holographic RG
flow equation \cite{dbvv},\cite{fms},\cite{mm}, in which the
temporal gauge is used for the metric in $(d+1)$ dimensions,
\begin{equation}
ds^2 = g_{ab}\, dx^a dx^b = dr^2 + g_{\mu\nu}(x,r)\, dx^\mu dx^\nu
\ , \qquad (\mu,\nu = 1, \cdots, d) \label{ads0}
\end{equation}
where $g_{\mu\nu}$ is dynamical in the bulk but supposed to tend
to the form of an ${\rm AdS}_{d+1}$ metric asymptotically at the
boundary $r \rightarrow \infty$ \cite{hs2},
\begin{equation}
g_{\mu\nu}(x,r) = e^{2r/l} \, \left[\, \hg_{\mu\nu}(x) -{l^2 \over
d-2}\left(\hR_{\mu\nu} - {1 \over 2(d-1)}\hR\,
\hg_{\mu\nu}\right)\, e^{-2r/l} + {\cal O}((e^{-2r/l})^2) \,
\right] \ . \label{ads1}
\end{equation}
The curvatures $\hR_{\mu\nu}$, $\hR$ are defined with the boundary
metric $\hg_{\mu\nu}(x)$, and $l$ is the ${\rm AdS}_{d+1}$ radius.
As the AdS metric diverges at the boundary, we will take the large
cut-off $r=r_0$. We consider the gravitational Lagrangian coupled
to a massive scalar field in the bulk,
\begin{equation}
{\cal L}_{d+1} = \km2 \sqrt{g} \, \left[- R_{d+1} + 2\Lambda  +
\hf g^{ab}\, \pd_a \phi \, \pd_b \phi+ \hf m^2 \phi^2 \right] \ ,
\label{Ls}
\end{equation}
where the cosmological constant $\Lambda=-d(d-1)/2l^2$. Then the
Hamiltonian reads
\begin{equation}
H = \int d^dx\, {\cal H} = \int d^dx\, \left[-\k2 g^{-1/2}\,
P_{\mu\nu\lm\sg}(g)\, \pi^{\mu\nu}\, \pi^{\lm\sg} - \hf \k2
g^{-1/2} \pi^2 + {\cal L}_d \right] \ ,
\end{equation}
where $P_{\mu\nu\lm\sg}(g) = \hf (g_{\mu\lm}\, g_{\nu\sg}+
g_{\mu\sg}\, g_{\nu\lm}) - {1 \over d-1}\, g_{\mu\nu}g_{\lm\sg}$
and
\begin{equation}
{\cal L}_d = \km2 \sqrt{g} \, \left[- R + 2\Lambda  + \hf
g^{\mu\nu}\, \pd_\mu \phi \, \pd_\nu \phi + \hf m^2 \phi^2 \right]
\ .
\end{equation}
Introducing the Hamilton-Jacobi functional $W(g,\phi)$ as
$\pi^{\mu\nu} = \dl W/\dl g_{\mu\nu}$ and $\pi = \dl W/\dl \phi$,
and inserting them into the Hamiltonian constraint ${\cal H}
\approx 0$, we obtain the holographic RG flow equation
\cite{dbvv},
\begin{equation}
{\cal H} =  - \{W,\, W\} + {\cal L}_d = 0 \ , \label{rgflow}
\end{equation}
where
\begin{equation}
\{W,\, W\} = \k2 g^{-1/2}\, \left[\left({\dl W \over \dl
g_{\mu\nu}}\right)^2 - {1 \over d-1}\, \left(g_{\mu\nu}{\dl W
\over \dl g_{\mu\nu}}\right)^2 + \hf \left({\dl W \over \dl
\phi}\right)^2 \right] \ .
\end{equation}
Note that the RG flow equation is defined on the surface $r=r_0$.
In $d=4$, we decompose the functional $W$ into the sum of
$S_{loc.}$ and $\Gm$,
\begin{equation}
S_{loc.} = \km2 \int d^4x \sqrt{g}\, \left[U(\phi) - \Phi(\phi)\,
R + \hf\, M(\phi)\, g^{\mu\nu}\pd_\mu \phi \pd_\nu \phi \right] \
, \label{sloc}
\end{equation}
where $U(\phi), \Phi(\phi), M(\phi)$ are functions of $\phi$
expanded as
\begin{eqnarray}
U(\phi) &=& U_0 + U_1\, \phi + \hf U_2\, \phi^2 + {\cal O}(\phi^3)
\ , \quad \Phi(\phi) = \Phi_0 + \Phi_1\, \phi + \hf \Phi_2\,
\phi^2 + {\cal O}(\phi^3) \ , \nonumber \\ \quad M(\phi) &=& M_0 +
{\cal O}(\phi) \ . \label{uphim}
\end{eqnarray}
The terms in $S_{loc.}$ are the first three terms appearing in the
derivative expansion of $W$. Inserting $W = S_{loc.} + \Gm$ into
(\ref{rgflow}) and comparing terms with the same weight $\om$ (the
number of differentiations) \cite{fms}, we have a series of
equations
\begin{eqnarray}\label{w02}
&&\om=0: \ - {U^2 \over 3} + \hf\, (U')^2 = - {12 \over l^2}
+\hf\, m^2 \phi^2 \ ,
\nonumber \\
&&\om=2: \ \left({U\Phi \over 3} - U' \Phi'\right) \, R - {ca
\over 6}\, g^{\mu\nu}\pd_\mu \phi \pd_\nu \phi +U'\left(-M\, \Box
\phi - \nabla^\mu M \nabla_\mu \phi + \hf M' g^{\mu\nu}\pd_\mu
\phi
\pd_\nu \phi \right) \nonumber \\
&&\hspace*{1.8cm} = -R + \hf \, g^{\mu\nu}\pd_\mu \phi \pd_\nu
\phi \ ,
\end{eqnarray}
where ${}'$ means the differentiation w.r.t. $\phi$, which leads
to
\begin{eqnarray}
U_0 &=& -6l^{-1} \qquad U_1 = 0 \qquad U_2 = l^{-1}\,
(\triangle_s-4) \nonumber \\
\Phi_0 &=& {l \over 2} \qquad \hspace{0.8cm}\Phi_1 = 0 \qquad
\Phi_2 = {l\,(\triangle_s-4) \over 12 (\triangle_s-3)} \qquad M_0
= {l \over 2 (\triangle_s-3)} \ , \label{coef}
\end{eqnarray}
where $\triangle_s = 2 + \sqrt{m^2l^2 + 4}$ is the scaling
dimension of a boundary operator associated with $\phi$, and
\begin{equation}
\om=4: \qquad \Phi_0^2 \, \km2 \sqrt{g}\,
\left(R^{\mu\nu}R_{\mu\nu} - {1 \over 3}R^2 \right) - {U_0 \over
3}\, g_{\mu\nu}{\dl \Gm \over \dl g_{\mu\nu}} + {\cal O}(\phi^2) =
0 \ . \label{w4}
\end{equation}
Finally, taking the boundary values of the metric
$g_{\mu\nu}(x,r_0) \rightarrow e^{2r_0/l}\hg_{\mu\nu}(x)$ and the
scalar field $\phi(x,r_0) \rightarrow 0$ as $r_0 \rightarrow
\infty$, and using the relations $\km2 = 1/16\pi G_5 =
Vol(S^5)/16\pi G_{10} = l^5\pi^3/16\pi(8\pi^6g_s^2l_s^8)$ and $l^8
= (4\pi)^2(g_s N)^2l_s^8$ in (\ref{w4}), we obtain the Weyl
anomaly of the boundary theory on the curved background
$\hg_{\mu\nu}$,
\begin{equation}
\langle T_\mu^{\ \mu}\rangle = -{2\over \sqrt{\hg}}\,
\hg_{\mu\nu}{\dl \Gm \over \dl \hg_{\mu\nu}} = {N^2 \over
32\pi^2}\, \left(\hR^{\mu\nu}\hR_{\mu\nu} - {1 \over 3}\hR^2
\right) \ , \label{weyl2}
\end{equation}
which exactly reproduces the Weyl anomaly of ${\cal N}=4$ $SU(N)$
SYM at leading (large $N$) order. Note that the scalar field does
not contribute to the final result (\ref{weyl2}) since, at leading
order, there only appear the first-order derivatives of $S_{loc.}$
w.r.t. $\phi$, which tend to zero when taking $\phi(x,r_0)
\rightarrow 0$ as $r_0 \rightarrow \infty$. However, at subleading
order, we have second-order derivatives w.r.t. the scalar field
(and also w.r.t. all the other Kaluza-Klein (KK) modes appearing
in the bulk supergravity \cite{krn}), which generally contribute
to the subleading Weyl anomaly even when they take their vanishing
boundary values, as will be shown below.
\par
\vspace{1.2cm}

As is well known, the exact form of the boundary Weyl anomaly is
given by (\ref{weyl2}) with the replacement $N^2 \rightarrow
N^2-1$. The factor $-1$ represents a $1/N^2$ correction to the
leading result and is expected to be derived from the quantum
(one-loop) calculation of the bulk supergravity. A generalized
version of the Hamiltonian constraint (\ref{rgflow}) responsible
for the quantum case is the Wheeler-DeWitt equation, which is a
quantum mechanical realization of (\ref{rgflow}) where a physical
state $\Psi$ has to be annihilated by the quantum operator $\cal
H$, which guarantees the `time' $r=r_0$ reparameterization
invariance of $\Psi$ as $\pd_{r_0} \Psi = -\int d^dx {\cal H}\,
\Psi = 0$. The physical state, when expressed in terms of path
integral, would be interpreted as the partition function with
boundary values, $\Psi(g,\phi)$,
\begin{equation}
{\cal H}\, \Psi = - \k2 g^{-1/2} \left( P_{\mu\nu\lm\sg}(g)\, {\dl
\over \dl g_{\mu\nu}}{\dl \over \dl g_{\lm\sg}} + \hf\, {\dl^2
\over \dl \phi^2} \right)\, \Psi + {\cal L}_d \, \Psi = 0 \ ,
\label{wdw1}
\end{equation}
or equivalently with $\Psi(g,\phi) = e^{-W(g,\phi)}$,
\begin{equation}
{\cal H}\, \Psi/\Psi = - \{W,\, W\} + {\cal L}_d + \k2 g^{-1/2}
\left( P_{\mu\nu\lm\sg}(g)\, {\dl^2 W \over \dl g_{\mu\nu} \dl
g_{\lm\sg}} + \hf\, {\dl^2 W \over \dl \phi^2} \right) = 0 \ ,
\label{wdw2}
\end{equation}
where the last two second-order derivative terms are of order $\k2
\sim N^{-2}$ and thus subleading corrections to (\ref{rgflow}). We
argue that this naive expression, however, leads to a misleading
result and have to be more careful to define the quantum
Hamiltonian operator associated with the time reparameterization.
\par

To explain  what is actually missing in (\ref{wdw2}), let us
consider the scalar part of the partition function $\Psi_s$, with
the boundary conditions, $\phi(x,r_i)=\phi_i(x)$ and
$\phi(x,r_f)=\phi_f(x)$,
\begin{equation}
\Psi_s = \int {\cal D}\phi \, e^{\int_{r_i}^{r_f}dr\int d^dx\,
{\cal L}_s(\phi, g)} \ , \qquad  {\cal L}_s = \km2 \sqrt{g} \,
\left[\, \hf \, g^{ab}\, \pd_a \phi \, \pd_b \phi+ \hf \, m^2
\phi^2 \right] \ ,
\end{equation}
where the path integral measure ${\cal D}\phi$ is induced by the
reparameterization-invariant inner product $||\dl \phi||^2 = \int
d^{d+1}x \sqrt{g}\, \dl\phi^2$. By the standard canonical path
integration, it is easily verified that $\Psi_s$ is expressed in
the operator formalism as
\begin{equation}
\langle \phi_f | \, T \exp\left[-\int_{r_i}^{r_f}dr\int d^dx\,
{\cal H}_s \right] \, | \phi_i \rangle = {\rm g}_f^{1/8} \ \Psi_s
\ {\rm g}_i^{1/8} \ , \qquad {\rm g}_{i(f)} = \prod_x g(x,
r_{i(f)}) \ .
\end{equation}
The extra factor ${\rm g}_f^{1/8}{\rm g}_i^{1/8}$ modifies the
ordinary Hamiltonian operator ${\cal H}_s$ into ${\tilde {\cal
H}}_s$ in the Schr\"{o}dinger equation, $\pd_{r_f}\, \Psi_s = -
\int d^dx\, {\tilde {\cal H}}_s \, \Psi_s$,
\begin{equation}
{\tilde {\cal H}}_s = \hf \km2 \sqrt{g_f}\, \left[- \kappa^4
g_f^{-1}\, {\dl^2 \over \dl \phi_f^2} + g_f^{\mu\nu}\pd_\mu \phi_f
\, \pd_\nu \phi_f + m^2 \phi_f^2 \right] + \pd_{r_f} {\rm ln}\,
g_f^{-1/8} \, \dl^d(0) \ . \label{sch}
\end{equation}
Note that the Schr\"{o}dinger equation is identical to that
derived in \cite{mn} when the metric is replaced with AdS
background metric. Comparing (\ref{wdw1}), (\ref{wdw2}) and
(\ref{sch}), we see that the extra subleading term $\pd_{r_0} {\rm
ln}\, g^{-1/8} \, \dl^d(0)$ is needed in the scalar part of the
Wheeler-DeWitt equation. It is straightforward to extend the
argument to higher spin cases such as vector, tensor and fermionic
fields \cite{kuy}; for example, a similar calculation for a
$(d+1)$-dimensional vector leads to its quantum Hamiltonian with
the extra term $\pd_{r_0} {\rm ln}\, g^{-(d-2)/8d} \, \dl_\mu^\mu
\dl^d(0)$ and so on. The Wheeler-DeWitt equation for the total
partition function $\Psi$ is thus defined with the subleading
$\dl^{d}(0)$ term for each KK particle appearing in ${\rm
AdS}_{d+1}$ supergravity.
\par

Let us estimate the subleading contribution to the boundary
anomaly arising from a five-dimensional scalar field, where the
extra term combined with the last term in (\ref{wdw2}) gives
\begin{eqnarray}
\hf \k2 g^{-1/2}{\dl^2 W \over \dl \phi^2} + \pd_{r_0} {\rm ln}\,
g^{-1/8} \, \dl^4(0) &=& \hf\, \left[M_0 (-\Box) - \Phi_2 R + U_2
+ \pd_{r_0} {\rm ln}\, g^{-1/4} \right] \, \dl^4(0) \nonumber \\
&+& \hf \k2 g^{-1/2}{\dl^2 \Gm \over \dl \phi^2} + {\cal O}(\phi)
\ ,
\end{eqnarray}
where we ignore the last two terms on the right-hand side (RHS)
since the second term is of weight $\om=8$, while $\phi$-dependent
terms vanish as $\phi \rightarrow 0$. The operator in the first
term stands in need of regularization; for example, it is carried
out by the zeta-function regularization in which the operator is
represented by a generalized zeta-function $\zeta(-1)$ as
described in detail in \cite{kuy}. After taking $r_0 \rightarrow
\infty$ and removing regularization-dependent divergent terms, we
have a finite term given by the heat-kernel coefficient $a_2(x,x)$
of $\om=4$ in the DeWitt-Schwinger proper time representation
\cite{bd}, showing that the subleading contribution does not
modify the result (\ref{coef}). In the operator, terms with
$-\Box$ and $R$ are of next-to-leading order $\sim {\cal
O}(e^{-2r_0/l})$ in the vicinity of the boundary, compared with
$U_2 \sim {\cal O}(1)$. We thus need leading and next-to-leading
terms of the asymptotic AdS metric (\ref{ads1}) to evaluate the
operator
\begin{equation}
\hf\, \left[M_0\left(-\Box + {1 \over 6}\, \hR \right) +
(\triangle_s -2)
 l^{-1} \right]\, \dl^4(0) =
{\sqrt{\hg} \over 32\pi^2}\, (\triangle_s -2)\, l^{-1} \,
a_2^{\xi=1/6}(x,x) \ , \label{sa2}
\end{equation}
which shows that a 5D minimally coupled scalar gives the
heat-kernel coefficient $a_2(x,x)$ of a 4D conformally coupled
scalar. As the RHS of (\ref{sa2}) comes in the RHS of the $\om=4$
equation (\ref{w4}), we have, at the boundary $r_0 \rightarrow
\infty$,
\begin{equation}
\langle T_\mu^{\ \mu}\rangle = {N^2 \over 32\pi^2}\,
\left(\hR^{\mu\nu}\hR_{\mu\nu} - {1 \over 3}\hR^2 \right) -
{\triangle_s -2 \over 32\pi^2}\, a_2^{\xi=1/6}(x,x) \ . \label{T}
\end{equation}
Interestingly, similar calculations for higher-spin fields in five
dimensions exhibit that their extra $\dl^4(0)$ terms lead to
four-dimensional conformally covariant operators, as the scalar
case \cite{kuy}. We see that the subleading correction (\ref{T})
for the scalar and those for the higher-spin fields to the leading
$N^2$ result are exactly the same as those previously obtained in
the Schr\"{o}dinger method \cite{mn2},\cite{mnu}, which guarantees
the desired shift $N^2 \rightarrow N^2 -1$ when summing up
contributions from all of the KK particles in AdS supergravity,
\begin{equation}
\langle T_\mu^{\ \mu}\rangle = {N^2 \over 32\pi^2}\,
\left(\hR^{\mu\nu}\hR_{\mu\nu} - {1 \over 3}\hR^2 \right) -
\sum_I{\triangle_I -2 \over 32\pi^2}\, a_2^I(x,x) = {N^2-1 \over
32\pi^2}\, \left(\hR^{\mu\nu}\hR_{\mu\nu} - {1 \over 3}\hR^2
\right) \ .
\end{equation}
\par

It is straightforward to generalize the above argument to $d={\rm
even}$ dimensional case in which $S_{loc.}$ in (\ref{sloc}) is
given by the sum of all possible local terms with weight $\om=0$
to $d-2$. For the leading large $N$ result, see
\cite{hs},\cite{fms}. In the massive scalar theory (\ref{Ls}), the
subleading correction is given by the heat-kernel coefficient
$a_{d/2}(x,x)$ for $d$-dimensional conformally coupled operator
$-\Box + \xi_d\, \hR$, with $\xi_d =(d-2)/4(d-1)$,
\begin{equation}
\langle T_\mu^{\ \mu}\rangle_{\rm subleading} = - {\triangle_s
-d/2 \over 2 (4\pi)^{d/2}}\, a_{d/2}^{\xi_d}(x,x) \ ,
\end{equation}
where $\triangle_s -d/2 = \sqrt{l^2m^2+(d/2)^2}$. It will be
discussed elsewhere how this result and those for higher-spin
fields contribute to the boundary Weyl anomaly at subleading order
in ${\rm AdS}_3/{\rm CFT}_2$ and ${\rm AdS}_7/{\rm CFT}_6$ cases.

\vspace{3.0cm}

The authors would like to thank Gabriele Veneziano for helpful
discussions at the early stage of the present work. The work of
T.K. is supported in part by Grant in Aid for Scientific Research
from the Japanese Ministry of Education (grant numbers 14540265
and 13135215). T.U. is supported by The Engineering and Physical
Sciences Research Council (EPSRC), UK. The work of N.Y. is
supported in part by the Japan Society for the Promotion of
Science and the Swiss National Science Foundation.

\pagebreak


\begin{thebibliography} {9}

\bibitem{m}{J.M. Maldacena, {\it The Large N Limit of Superconformal Field
Theories and Supergravity}, Adv.~Theor.~Math.~Phys.~2 (1998) 231;
Int.~J.~Theor.~Phys.~38 (1999) 1113.}

\bibitem{hs}{M. Henningson and K. Skenderis, {\it The holographic
Weyl anomaly}, JHEP~9807 (1998) 023.}

\bibitem{no}{S. Nojiri and S.D. Odintsov, {\it Conformal anomaly for
dilaton coupled theories from AdS/CFT correspondence},
Phys.~Lett.~B444 (1998) 92; S. Nojiri and S.D. Odintsov, {\it On
the Conformal Anomaly from Higher Derivative Gravity in AdS/CFT
Correspondence}, Int.~J.~Mod.~Phys.~A15 (2000) 413; S. Nojiri,
S.D. Odintsov and S. Ogushi, {\it Finite action in 5D gauged
supergravity and the dilatonic conformal anomaly for dual quantum
field theory}, Phys.~Rev.~D62 (2000) 124002.}

\bibitem{mn}{P. Mansfield and D. Nolland, {\it One-loop conformal anomalies
from AdS/CFT in the Schr\"{o}dinger representation}, JHEP~9907
(1999) 028.}

\bibitem{mn2}{P. Mansfield and D. Nolland, {\it Order $1/N^2$ test of the
Maldacena conjecture: cancellation of the one-loop Weyl anomaly},
Phys.~Lett.~B495 (2000) 435.}

\bibitem{mnu}{P. Mansfield, D. Nolland and T. Ueno, {\it Order $1/N^2$ test
of the Maldacena conjecture II: the full bulk one-loop
contribution to the boundary Weyl anomaly}, Phys.~Lett.~B565
(2003) 207.}

\bibitem{dbvv}{J. De Boer, E. Verlinde and H. Verlinde, {\it On
the holographic renormalization group}, JHEP~0008 (2000) 003.}

\bibitem{fms}{M. Fukuma, S. Matsuura and T. Sakai, {\it A Note on
the Weyl Anomaly in the Holographic Renormalization Group},
Prog.~Theor.~Phys.~104 (2000) 1089; M. Fukuma, S. Matsuura and T.
Sakai, {\it Holographic Renormalization Group},
Prog.~Theor.~Phys.~109 (2003) 489.}

\bibitem{mm}{D. Martelli and W. M\"{u}ck, {\it Holographic renormalization and
Ward identities with the Hamilton-Jacobi method}, Nucl.~Phys.~B654
(2003) 248.}

\bibitem{hs2}{M. Henningson and K. Skenderis, {\it Holography and the
Weyl anomaly}, Fortsch.~Phys.~48 (2000) 125, hep-th/9812032.}

\bibitem{krn}{H.J. Kim, L.J. Romans and P. van Nieuwenhuizen,
{\it Mass spectrum of chiral ten-dimensional $N=2$ supergravity on
$S^5$}, Phys.~Rev.~D32 (1985) 389.}

\bibitem{kuy}{T. Kubota, T. Ueno and N. Yokoi, in preparation.}

\bibitem{bd}{For example, N.D. Birrell and P.C. Davies,
{\it Quantum fields in curved space} (Cambridge University Press,
1982).}

\end{thebibliography}
\end{document}